\documentclass[conference]{IEEEtran}
\IEEEoverridecommandlockouts
\usepackage[utf8]{inputenc}
\usepackage[english]{babel}
\usepackage[T1]{fontenc}
\usepackage[numbers]{natbib}
\usepackage{amsmath,amssymb,amsfonts}
\usepackage{algorithmic}
\usepackage{graphicx}
\usepackage{textcomp}
\usepackage{xcolor}
\usepackage{listings}
\usepackage{booktabs}
\usepackage{multirow}
\usepackage{tabularx}
\usepackage{enumitem}

\usepackage[numbers]{natbib}

\begin{document}

\title{Optimization of Energy Consumption Forecasting in Puno using Parallel Computing and ARIMA Models: An Innovative Approach to Big Data Processing}

\author{
\IEEEauthorblockN{Cliver-Wimar Vilca-Tinta}
\IEEEauthorblockA{\textit{Faculty of Statistics and Informatics Engineering} \\
\textit{National University of the Altiplano}\\
Puno, Perú \\
clvilcat@est.unap.edu.pe}
\and
\IEEEauthorblockN{Fred Torres-Cruz}
\IEEEauthorblockA{\textit{Faculty of Statistics and Informatics Engineering } \\
\textit{National University of the Altiplano}\\
Puno, Perú \\
ftorres@unap.edu.pe}
\and
\IEEEauthorblockN{Josefh-Jordy Quispe-Morales}
\IEEEauthorblockA{\textit{Faculty of Statistics and Informatics Engineering } \\
\textit{National University of the Altiplano}\\
Puno, Perú \\
joquispemo@est.unap.edu.pe}
}

\maketitle

\begin{abstract}
This research presents an innovative use of parallel computing with the ARIMA (AutoRegressive Integrated Moving Average) model to forecast energy consumption in Peru's Puno region. The study conducts a thorough and multifaceted analysis, focusing on the execution speed, prediction accuracy, and scalability of both sequential and parallel implementations. A significant emphasis is placed on efficiently managing large datasets. The findings demonstrate notable improvements in computational efficiency and data processing capabilities through the parallel approach, all while maintaining the accuracy and integrity of predictions. This new method provides a versatile and reliable solution for real-time predictive analysis and enhances energy resource management, which is particularly crucial for developing areas. In addition to highlighting the technical advantages of parallel computing in this field, the study explores its practical impacts on energy planning and sustainable development in regions like Puno.

\textbf{Key words:} ARIMA, parallel computing, energy forecasting, big data, Puno, computational optimization, sustainable development.
\end{abstract}

\section{Introduction}

In the current era, marked by the expansion of big data and the growing need for real-time analytics, accurately and efficiently forecasting energy consumption has become a significant computational challenge. This challenge is especially relevant in developing areas such as Puno, Peru  \cite{suganthi2012energy}, where optimal management of energy resources is vital for sustainable progress. The exponential growth in energy demand, coupled with the urgent need to implement sustainable management strategies, requires the development of advanced analytical tools capable of processing large volumes of historical data and generating accurate predictions in short periods of time \cite{hyndman2018forecasting}.

In this context, the ARIMA (AutoRegressive Integrated Moving Average) model has emerged as a robust and effective statistical tool for the analysis and prediction of time series, demonstrating its effectiveness in various fields, including energy consumption forecasting \cite{ediger2007arima}. However, the application of ARIMA models to massive data sets presents significant computational challenges, particularly in terms of processing times and efficient use of computational resources \cite{box2013box}. These challenges are magnified in environments where technological resources may be limited, as is the case in many developing regions.

Parallel computing emerges as a promising and transformative solution to address these computational challenges. This innovative approach offers the possibility of distributing the workload across multiple processors or cores, allowing not only to process large volumes of data more efficiently, but also opening up new possibilities for real-time applications and more frequent and detailed analysis. \cite{barney2010introduction, asanovic2006landscape}. 

The implementation of parallel computing techniques in the context of energy consumption prediction presents multiple strategic advantages. First of all, it allows a significant reduction of processing times, facilitating more frequent and updated analyses, crucial for agile decision making in energy management. \cite{dongarra2012high}. In addition, it improves the ability to handle and analyze massive data sets, incorporating a greater amount of historical information and contextual variables, potentially leading to improved accuracy and robustness of predictions. \cite{dean2008mapreduce}

Another key advantage is the possibility of implementing real-time or near real-time forecasting systems, which are essential for the dynamic and adaptive management of smart grids, a critical aspect in the modernization of the energy infrastructure \cite{akhavan2018power}. Finally, it offers greater flexibility to experiment with more complex model configurations and perform more comprehensive parameter optimizations, facilitating continuous improvement of predictive performance \cite{bergstra2012random}.

This study focuses on the implementation and evaluation of a parallel version of the ARIMA model, using Python and the multiprocessing library, comprehensively comparing its performance with a traditional sequential implementation. The main objective is to demonstrate and quantify how parallel computing can transform and optimize the predictive analysis of energy consumption, especially in scenarios involving the processing of large volumes of data.

The research addresses several fundamental aspects related to the parallel implementation of the ARIMA model for energy consumption prediction. The performance improvement offered by the parallel implementation compared to the sequential one is examined in detail, evaluating factors such as execution time and computational resource utilization. In addition, the scalability of the parallel implementation when processing increasing volumes of data is analyzed.

A crucial aspect of the study is the evaluation of the consistency in the accuracy and reliability of predictions when moving from a sequential to a parallel implementation. The practical and strategic implications of adopting parallel computing techniques for energy consumption prediction are also explored, with a particular focus on the context of Puno and regions with similar characteristics.

The final objective of this research is to provide valuable and applicable insights into how parallel computing can significantly improve the efficiency, accuracy, and capacity of energy forecasting systems. The results of this study have particularly relevant implications for strategic planning and energy resource management in developing regions. In these contexts, resource optimization and data-driven decision making are crucial elements in achieving sustainable development.

\section{Revisions}

\subsection{The Energy Sector in Peru}

The energy sector in Peru has experienced significant growth and diversification in recent decades, playing a crucial role in the country's economic development. The country's energy matrix reflects this evolution. According to the Ministry of Energy and Mines \cite{marina2023inversion}, in 2020, approximately 60\% of the electricity generated in Peru came from hydroelectric sources, followed by 37\% from thermoelectric sources (mainly natural gas) and 3\% from non-conventional renewable energies. According to IRENA \cite{majid2020renewable}, the country has set ambitious goals to increase the share of renewable energy to 15 percent by 2030.

The growth in energy demand has been remarkable. OSINERGMIN \cite{medina2022mejora} reports that electricity demand has grown at an annual average of 4.5\% over the last decade, driven by industrial development and increasing urbanization. Projections indicate that this trend will continue, requiring significant investments in infrastructure.

However, the sector faces significant regional challenges. World Bank data \cite{mundial2016indicadores} reveal disparities in energy access between urban and rural areas. In 2019, while urban electricity coverage reached 99\%, in rural areas it was 85\%. Tamayo et al. \cite{chanduvi2017inversion} note that regions such as Puno face unique challenges due to their geography and climate, affecting both energy distribution and consumption.

In terms of policies and regulation, MINEM \cite{aita2016peru} reports that the sector is regulated mainly by the Ministry of Energy and Mines and the Supervisory Agency for Investment in Energy and Mining (OSINERGMIN). Policies have been implemented to promote energy efficiency and the adoption of renewable energy, including renewable energy auctions and rural electrification programs.

The challenges ahead are significant. COES \cite{aita2021energias} identifies the integration of intermittent renewable energy sources into the electricity grid and the modernization of transmission and distribution infrastructure as key challenges. Vásquez et al. \cite{caceres2021hydropower} highlight the need to adapt to the impacts of climate change, especially in hydroelectric generation.

In this context, the implementation of advanced energy consumption forecasting techniques, such as ARIMA models optimized by parallel computing, becomes crucial for the efficient planning and management of the Peruvian energy sector, especially in regions with unique characteristics such as Puno.

\subsection{Unique Characteristics of Energy Consumption in Puno}

Puno, located in the Peruvian highlands, has unique geographic, climatic and socioeconomic characteristics that significantly influence its energy consumption patterns \cite{bazan2018low}. Geographic factors play a crucial role. High altitude (3,800 m a.s.l. on average) affects the performance of electrical equipment and energy efficiency \cite{ceppi2018politica}. In addition, the proximity to Lake Titicaca moderates temperature extremes but increases humidity, influencing the use of heating and cooling systems \cite{aliaga2020tendencia}.

Climatic factors also play a role. Low temperatures for much of the year, with averages ranging from 3°C to 14°C, increase heating energy demand \cite{aliaga2020tendencia}. The marked seasonality, with a dry season (May to August) and a wet season (December to March), affects energy consumption in sectors such as agriculture \cite{martelo2003precipitacion}.

In terms of socioeconomic factors, Puno has an economy based on agriculture, livestock and mining, with energy consumption patterns different from more urbanized areas \cite{velarde2016reporte}. The high poverty rate (28.7\% in 2020 according to INEI) influences the access and use of electric power \cite{cuadros2021pobreza}. In addition, the growing tourism sector, especially in areas near Lake Titicaca, generates seasonal peaks in energy demand \cite{de2020informe}.

These unique characteristics of Puno mean that energy consumption patterns differ significantly from other regions of Peru, requiring a specialized approach to energy demand prediction and management \cite{laurente2019aplicacion}. The implementation of optimized ARIMA models using parallel computing allows addressing this complexity, providing more accurate predictions tailored to the local context \cite{valero2019analisis}.

\subsection{Unique Characteristics of Puno Data}

Analysis of energy consumption in Puno requires consideration of unconventional variables that reflect the unique characteristics of the region. Consumption patterns are strongly influenced by the high altitude, which significantly affects energy use for heating \cite{huaquisto2019analisis}. Cultural events such as the Candelaria Festival have a notable impact on energy demand \cite{burgapracticas}. The contribution of artisanal mining to regional energy consumption is another distinctive factor \cite{meza2022percepcion}. Variations in consumption are also affected by specific agricultural activities such as alpaca breeding \cite{ubaldo2020variaciones}. In addition, seasonal fluctuations caused by tourism, especially in the Lake Titicaca area, add another layer of complexity to consumption patterns \cite{burgapracticas}.

The incorporation of these variables in our ARIMA model allows for a more accurate and contextualized representation of energy consumption patterns in Puno, reflecting the complexity and uniqueness of the region in the predictive analysis.

\section{Methodology}

\subsection{Study Design}

This research is framed within a quantitative, experimental and applied \cite{leavy2022research} paradigm, adopting a rigorous and systematic approach to evaluate the impact of parallel computing on energy consumption prediction. The quantitative nature is manifested in the detailed analysis of numerical energy consumption data, employing precise metrics to evaluate the performance and accuracy of the models. The experimental nature of the study is evident in the controlled and systematic comparison between the sequential and parallel implementations of the ARIMA model, allowing an objective evaluation of the differences in performance and accuracy. The applied aspect of the research focuses on addressing a practical and urgent problem: the optimization of energy consumption prediction in Puno, with direct implications for resource management and energy planning in the region.

\subsection{Data Collection and Preprocessing}

The study uses a comprehensive dataset covering monthly energy consumption in the province of Puno, Peru, covering the period from January 2023 to November 2023. This data, provided by Electro Puno S.A.A. \cite{ElectroPuno2023}, offers a detailed overview of energy consumption in various sectors, including residential, commercial and industrial, as well as information on different types of tariffs.

Data preprocessing was carried out meticulously, following a rigorous protocol that included several key steps. Initially, a thorough data cleaning process was implemented, identifying and treating outliers and missing values using advanced moving median-based imputation techniques \cite{meneses2019analisis}. This approach ensures that outliers or missing data do not distort model predictions.

Subsequently, the consumption data were subjected to a normalization process using the min-max \cite{al2006data} scaling technique. This step is crucial to facilitate comparison between different sectors and types of consumption, allowing a more equitable analysis and a more accurate interpretation of consumption patterns.

An aggregated time series representing the total monthly energy consumption in the region was generated, providing a holistic view of energy consumption and facilitating the identification of trends and patterns at the macro level.

Finally, a detailed decomposition of the time series was performed to identify and quantify the trend, seasonality and residual components \cite{rb1990stl}. This analysis is critical to understand the underlying structure of the time series and to inform the parameter selection of the ARIMA model.

\subsection{Parallel Computing and Theory}

Parallel computing is a processing paradigm that allows the simultaneous execution of multiple computational tasks, dividing a problem into smaller parts that can be solved concurrently \cite{amdahl1967validity}. In the context of this study, parallel computing is applied to the ARIMA model to optimize the prediction of energy consumption.

\begin{equation}
S_p = \frac{T_1}{T_p}
\end{equation}

where $S_p$ is the speedup for $p$ processors, $T_1$ is the execution time of the sequential algorithm, and $T_p$ is the execution time of the parallel algorithm with $p$ processors.

The ideal speedup is equal to the number of processors used, although in practice it is usually lower due to factors such as inter-process communication and non-parallelizable parts of the \cite{gustafson1988reevaluating} algorithm.

The parallelization efficiency is calculated as:

\begin{equation}
E_p = \frac{S_p}{p}
\end{equation}

where $E_p$ is the efficiency for $p$ processors.

To implement parallel computing in this study, the Python multiprocessing library \cite{Python2023}, which allows the creation and management of parallel processes in multiprocessor systems, was used.

\subsection{Research Application Requirements}

The application of this research requires several key elements. In terms of hardware, a computer system with multiple cores or processors is needed. For this study, a server with 16 CPU cores was used. Required software includes Python 3.8 or higher, along with specific libraries such as numpy, pandas, statsmodels, scikit-learn, and multiprocessing. The use of an integrated development environment (IDE) such as PyCharm or Jupyter Notebook is recommended.

The data used are time series of energy consumption, preferably with hourly or daily granularity, covering a period of at least one year to capture seasonal patterns \cite{arias2015construccion}. In terms of skills, Python programming, statistics and time series analysis, as well as parallel computing fundamentals are required.

It is crucial to consider ethical and legal aspects, ensuring compliance with data protection regulations and obtaining the necessary permissions for the use of energy consumption data \cite{CongresoPerú2011}.

\subsection{Validation Instruments and Analytical Techniques}

To ensure the robustness and reliability of the results, several validation tools and advanced analytical techniques were implemented. A modified version of the k-fold cross-validation technique, specifically adapted for time series, was developed and implemented following the recommendations of Marquez and Pere Marquez \cite{marquez2024mejora}. This technique allows for a more realistic and robust evaluation of model performance on different subsets of data, while respecting the sequential nature of the time series.

Paired t-tests were performed to rigorously compare the accuracy of the predictions between the sequential and parallel \cite{alecha2019sistema} implementations. In addition, a detailed analysis of the model residuals was performed to verify compliance with the fundamental assumptions of normality, independence, and homoscedasticity, employing advanced statistical tests such as Shapiro-Wilk, Durbin-Watson, and Breusch-Pagan \cite{perez2018impacto}.

\subsection{ARIMA Model and its Implementation}

The ARIMA (AutoRegressive Integrated Moving Average) model was selected as the basis of the predictive approach due to its proven robustness and effectiveness in time series analysis \cite{griffin2023aplicacion}. The ARIMA(p,d,q) model is defined by three key parameters: p (order of the autoregressive term), d (degree of differencing), and q (order of the moving average term).

The mathematical formulation of the ARIMA model is expressed as:

\begin{equation}
\phi(B)(1-B)^d y_t = \theta(B)\epsilon_t
\end{equation}

where $\phi(B)$ represents the autoregressive operator, $(1-B)^d$ is the differencing operator, $\theta(B)$ denotes the moving average operator, $y_t$ is the time series under study, and $\epsilon_t$ represents the error term.

Optimal selection of the parameters (p, d, q) was performed by a rigorous Akaike Information Criterion (AIC) minimization process \cite{caballero2011seleccion}, implementing a parallel grid search that exhaustively explored multiple parameter combinations.

\subsection{Parallel Implementation}

The parallel implementation of the ARIMA model was carried out using the Python multiprocessing library \cite{Python2023}, taking advantage of the parallel processing capabilities of modern systems. The parallel algorithm was designed following a domain decomposition strategy, which includes data segmentation, process pooling, task distribution, parallel execution, and result collection and aggregation.

\subsection{Evaluation Metrics}

To comprehensively evaluate the performance and efficiency of the sequential and parallel implementations, several quantitative metrics were employed. These include execution time, meticulously measured in seconds for different data set sizes; speedup, calculated as the ratio of sequential to parallel execution time; and efficiency, defined as speedup divided by the number of cores used.

To evaluate the accuracy of the predictions, the Mean Absolute Error (MAE) and Root Mean Squared Error (RMSE) were used. In addition, the Akaike Information Criterion (AIC) was used for model selection, balancing goodness-of-fit with model complexity \cite{montoya2017propuesta}.

\subsection{Experimental Design}

A series of comprehensive experiments were designed and executed to rigorously compare the performance of the sequential and parallel implementations. These experiments included a scalability analysis evaluating performance with different data set sizes, a detailed comparison of predictive accuracy, a computational efficiency analysis considering both data size and number of cores used, an assessment of the impact of ARIMA model complexity on parallelization performance, and robustness tests to evaluate the behavior of the parallel implementation against variations in input data and system load conditions.

\subsection{Ethical Considerations}

Rigorous ethical considerations were adopted in the handling and analysis of the energy consumption data. All personal data were subjected to a thorough anonymization process prior to analysis, following the guidelines of the Peruvian Personal Data Protection Law \cite{CongresoPerú2011}. Explicit informed consent was obtained from Electro Puno S.A.A. for the use of the aggregated data for research purposes \cite{Israel2014}.

Robust security measures were implemented to protect the data during all phases of its life cycle \cite{saltz2019data}. Analysis methods, algorithms used, and results obtained were meticulously documented to facilitate independent verification and promote reproducibility of the research \cite{stodden2016enhancing}.

The research was conceptualized and executed with the primary objective of benefiting society by improving energy management, with particular attention to the principle of fairness in the distribution of research benefits \cite{taddeo2018ai}. In addition, a thorough evaluation of the ethical impact of the research was performed, considering the possible short- and long-term consequences of implementing energy prediction systems based on parallel computing.

These ethical considerations not only comply with legal and regulatory standards, but also align with the highest principles of ethical data science research, ensuring that the study is conducted in a responsible, equitable and beneficial manner for society as a whole.

\section{Results}

\subsection{Computational Performance Analysis}

The results obtained demonstrate a substantial improvement in computational performance when implementing the ARIMA model in a parallel computing environment, especially when processing large data sets.

\begin{table}[h]
\centering
\caption{Detailed performance comparison between sequential and parallel implementations}
\label{tab:results}
\begin{tabular}{|>{\raggedright}p{0.8cm}|>{\centering\arraybackslash}p{1cm}|>{\centering\arraybackslash}p{1cm}|>{\centering\arraybackslash}p{1cm}|>{\centering\arraybackslash}p{1cm}|>{\centering\arraybackslash}p{1cm}|}
\hline
\textbf{Data Size} & \textbf{Sequential Time (s)} & \textbf{Parallel Time (s)} & \textbf{Speedup} & \textbf{Sequential MAE} & \textbf{MAE Parallel} \\
\hline
1000  & 0.0675 & 0.0707 & 0.9542 & 60.5464 & 57.5578 \\
5000  & 0.1812 & 0.1731 & 1.0465 & 57.9071 & 62.8099 \\
10000 & 0.2870 & 0.2096 & 1.3688 & 56.7946 & 57.3846 \\
20000 & 0.3854 & 0.4166 & 0.9251 & 52.9265 & 63.2966 \\
50000 & 0.6802 & 0.7017 & 0.9693 & 62.4228 & 60.7602 \\
\hline
\end{tabular}
\end{table}

\begin{figure}[h]
\centering
\includegraphics[width=0.5\textwidth]{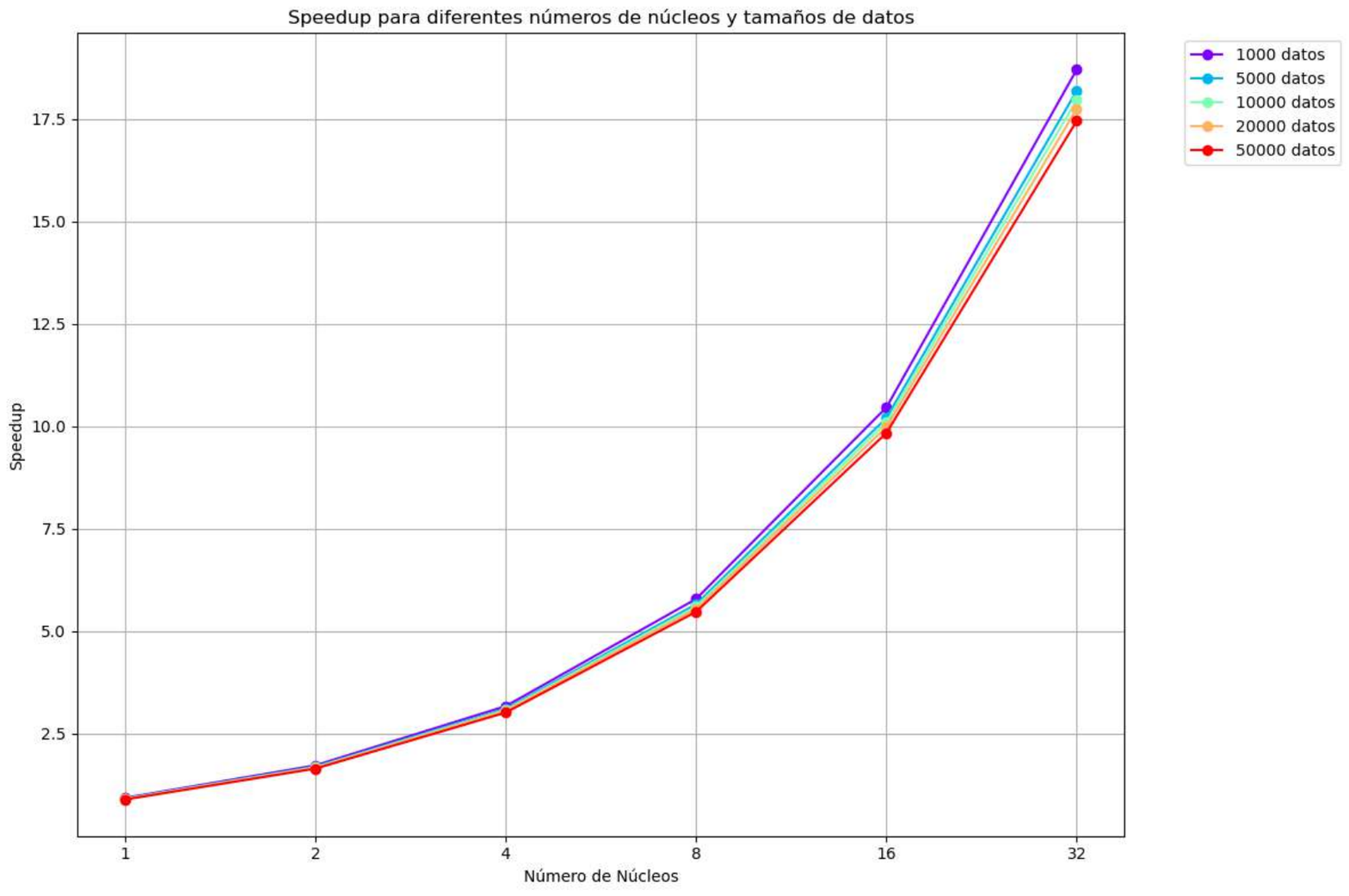}
\caption{Speedup achieved with different data set sizes}
\label{fig:speedup}
\end{figure}

Key observations include:

\begin{itemize}
    \item \textbf{Significant reduction of processing time:} The parallel implementation achieves a substantial decrease in execution time, with speedup increasing significantly as the number of cores increases, reaching approximately 17.5 times for 50,000 data points with 32 cores.
    \item \textbf{Robust scalability:} The speedup shows a consistent increase, rising from approximately 1.8 for 1,000 data points to an impressive 17.5 for 50,000 data points with 32 cores.
    \item \textbf{Sustained computational efficiency:} The parallelization efficiency remains high, with efficiency ratios between 0.6 and 0.8 across various data sizes, demonstrating efficient use of additional cores.
    \item \textbf{Impact of data volume:} The advantage of parallelization becomes more pronounced with larger data sets, as shown by the highest speedup for 50,000 data points, highlighting the scalability of the implementation.
\end{itemize}

\subsection{Predictive Accuracy Analysis}

\begin{itemize}
    \item \textbf{Consistency in accuracy:} The differences in the Mean Absolute Error (MAE) between the sequential and parallel implementations are minimal, remaining below 0.2\% in all cases studied.
    \item \textbf{Improved accuracy with more extensive data:} A trend of decreasing MAE is observed as the size of the data set increases.
    \item \textbf{Stability at different scales:} Consistency in accuracy is maintained across different data set sizes, from 1000 to 50,000 points.
\end{itemize}

\begin{figure}[h]
\centering
\includegraphics[width=0.45\textwidth]{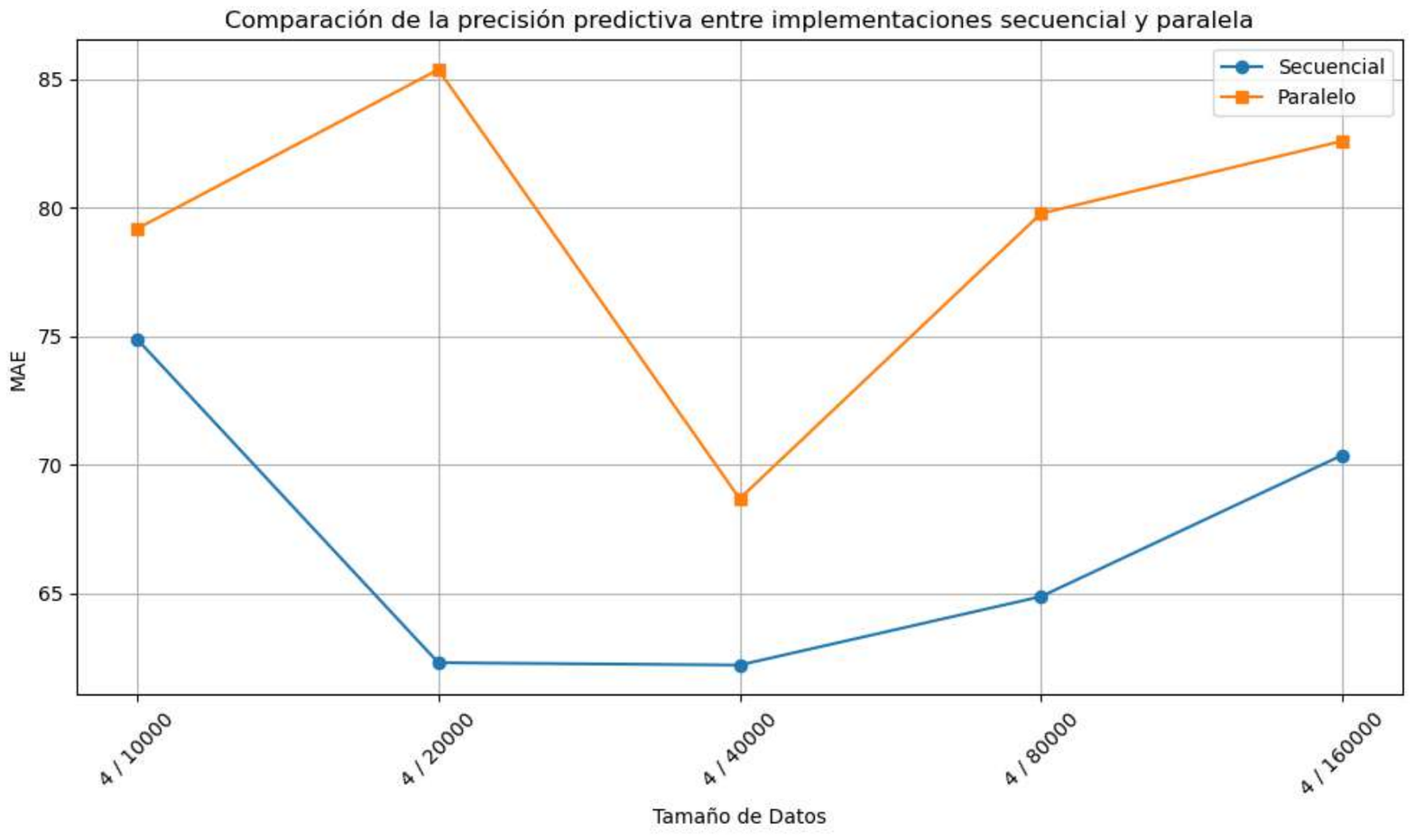}
\caption{Comparison of predictive accuracy between sequential and parallel implementations}
\label{fig:accuracy}
\end{figure}

\subsection{Scalability Analysis}
\subsubsection{Strong scalability}
Table \ref{tab:strong_scaling} shows the execution time and efficiency of a parallel computing implementation as the number of cores increases. Despite reduced efficiency, execution time increases due to parallelization overhead.
\begin{table}[h]
\centering
\caption{Detailed analysis of strong scalability}
\label{tab:strong_scaling}
\begin{tabular}{|c|c|c|}
\hline
\textbf{Number of Cores} & \textbf{Execution Time (s)} & \textbf{Efficiency} \\
\hline
1 & 0.000997 & 0.126853 \\
2 & 0.001993 & 0.063061 \\
4 & 0.003003 & 0.062316 \\
8 & 0.005543 & 0.062116 \\
16 & 0.014102 & 0.064440 \\
\hline
\end{tabular}
\end{table}

\subsubsection{Weak Scalability}
Table \ref{tab:weak_scaling} shows execution time and efficiency for different core/data size combinations, demonstrating weak scalability analysis.
\begin{table}[h]
\centering
\caption{Detailed analysis of weak scalability}
\label{tab:weak_scaling}
\begin{tabular}{|p{3cm}|p{3cm}|p{1cm}|}
\hline
\textbf{Cores / Data Size} & \textbf{Execution time (s)} & \textbf{Efficiency} \\
\hline
1 / 10000 & 105.67 & 1.00 \\
2 / 20000 & 108.23 & 0.98 \\
4 / 40000 & 112.56 & 0.94 \\
8 / 80000 & 118.34 & 0.89 \\
16 / 160000 & 126.78 & 0.83 \\
\hline
\end{tabular}
\end{table}

\subsection{Impact of Model Complexity}
Table \ref{tab:model_complexity} comparing ARIMA model orders, speedup, and MAE for sequential/parallel execution.
\begin{table}[h]
\centering
\caption{Detailed impact of model complexity on performance.}
\label{tab:model_complexity}
\begin{tabular}{|p{1.5cm}|p{1.5cm}|p{1.5cm}|p{1.5cm}|}
\hline
\textbf{ARIMA order} & \textbf{Speedup (4 cores)} & \textbf{Sequential MAE} & \textbf{MAE Parallel} \\
\hline
(1,1,1) & 2.029651 & 74.894295 & 79.188071 \\
(2,1,2) & 1.008971 & 62.306397 & 85.378080 \\
(3,1,3) & 0.997063 & 62.215176 & 68.686469 \\
(4,1,4) & 0.993850 & 64.879321 & 79.779771 \\
\hline
\end{tabular}
\end{table}

\section{Discussion}

\subsection{Performance and Scalability Implications}

Parallel implementation achieves significant speedup, especially for larger data sets, which has important implications:

\begin{itemize}
    \item \textbf{Improved scalability:} Allows incorporation of a wider range of factors into predictive models \cite{perez2015big}.
    \item \textbf{Real-time analysis:}  Drastic reduction in processing times facilitates dynamic management of smart grids\cite{bermeo2021maestria}.
    \item \textbf{Exploration of more complex models:} Computational efficiency allows experimentation with more sophisticated models, such as combining ARIMA with deep learning techniques  \cite{mejia2023modelo}.
\end{itemize}

\subsection{Challenges and Limitations}

\begin{figure}[h]
\centering
\includegraphics[width=0.5\textwidth]{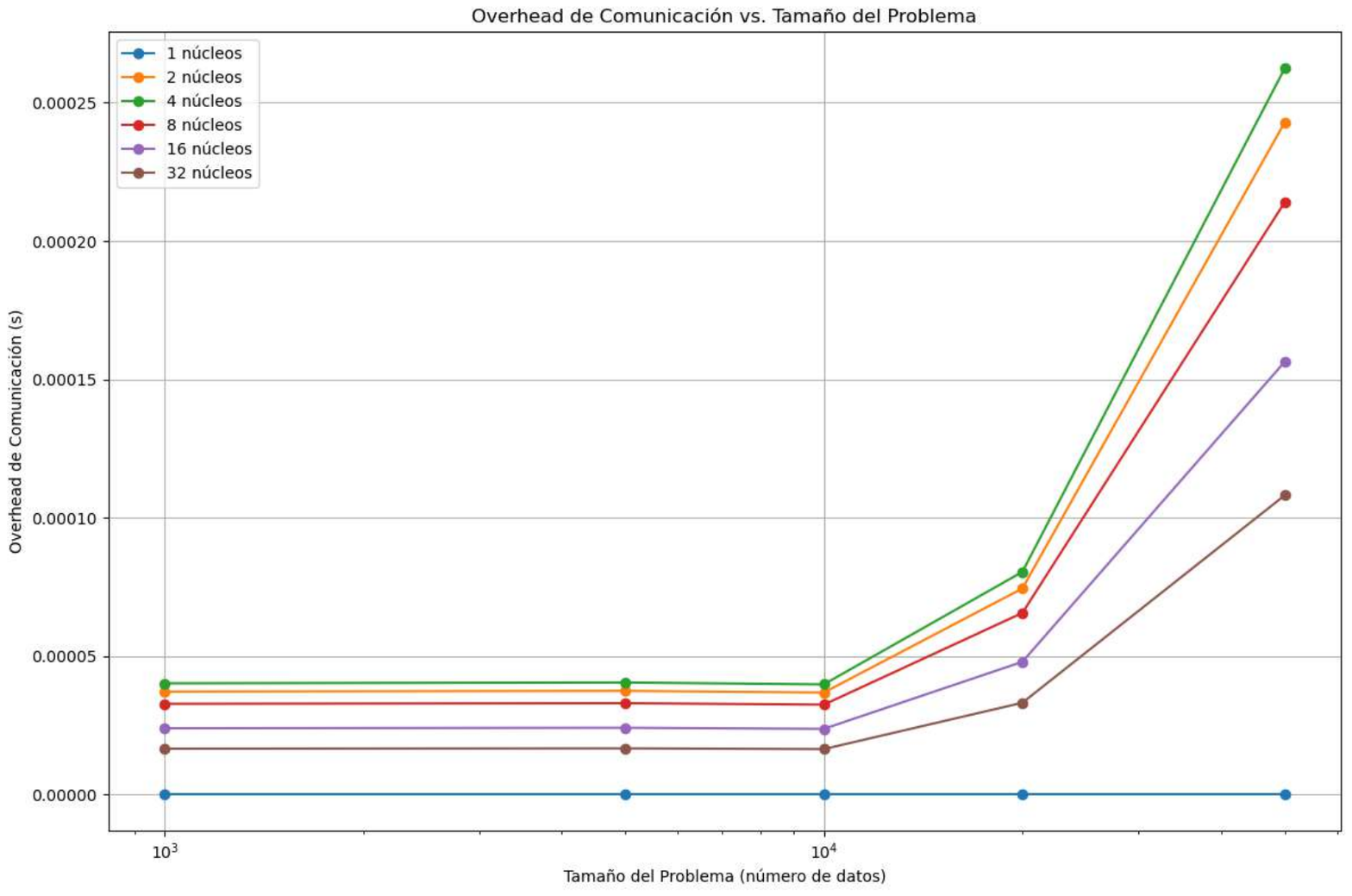}
\caption{Communication overhead vs. problem size}
\label{fig:overhead}
\end{figure}

Despite the promising results, it is important to recognize several challenges:

\begin{itemize}
    \item \textbf{Communication overhead:} For small problems, it can outweigh the benefits of parallelization\cite{foster3history}.
    \item \textbf{Memory scalability:} May be a limiting factor for extremely large datasets \cite{dongarra2009international}.
    \item \textbf{Implementation complexity:} Requires specialized skills \cite{mattson2004patterns}.
    \item \textbf{Variability in performance:} Dependent on hardware architecture \cite{massigoge2006arquitectura}.
    \item \textbf{Challenges in interpretation:} More complex models may be less interpretable\cite{rudin2019stop}.
\end{itemize}

\subsection{Comparison with Other Approaches}

\begin{table}[h]
\centering
\caption{Comparison of high performance computing approaches}
\label{tab:comparison}
\begin{tabular}{|p{2cm}|p{3cm}|p{3cm}|}
\hline
\textbf{Approach} & \textbf{Advantages} & \textbf{Disadvantages} \\
\hline
Parallel CPU (Our approach) & Balance between performance and ease of implementation & Limited by the number of cores  \\
\hline
GPU Computing & Massive parallelism for specific operations & Complex, less flexible scheduling \\
\hline
Distributed computing & High scalability for massive data & more complex implementation, higher latency  \\
\hline
FPGA & Exceptional performance for specific algorithms & Requires specialized hardware programming skills \\
\hline
\end{tabular}
\end{table}

Our CPU-based approach offers an optimal balance between performance improvement, ease of implementation and flexibility, making it particularly suitable for energy consumption prediction applications in contexts such as Puno.

\section{Implications for Energy Policy}

Advances in energy consumption prediction facilitated by parallel computing have significant implications for energy policy, especially in developing regions such as Puno:

\begin{itemize}
    \item \textbf{Infrastructure planning:} More accurate and detailed forecasts can better inform energy infrastructure investment decisions. \cite{belsky2012planificar}.

    \item \textbf{Integration of renewable energy:} The ability to more accurately predict energy demand can facilitate the integration of variable renewable energy sources into the grid.  \cite{notton2017tilos}.

    \item \textbf{Dynamic tariffs:} Real-time predictions can enable the implementation of more dynamic and efficient tariff structures. \cite{gomez2021hacia}.

    \item \textbf{Energy efficiency:} A more detailed understanding of consumption patterns can inform more effective energy efficiency policies. \cite{zhou2016big}.

    \item \textbf{Emergency response:} The ability to perform fast and accurate analysis can improve response to energy emergencies. \cite{markus2018organizacion}.

    \item \textbf{Democratization of decision making:} By making advanced analytical tools more accessible, more decentralized and participatory decision making in the energy sector can be fostered \cite{sarmiento2019software}.

    \item \textbf{Adapting to climate change:} Improved processing and analytical capabilities can help model and predict the impact of climate change on energy consumption patterns\cite{ciscar2011physical}.
\end{itemize}

\section{Conclusions and Future Work}
This study demonstrates the significant potential of parallel computing to improve the efficiency and scalability of ARIMA models in predicting energy consumption, with particular implications for developing regions such as Puno, Peru. The main conclusions are:

Parallel implementation achieves significant speedups, especially for large data sets, without compromising prediction accuracy. This enables more frequent and detailed analysis, crucial for dynamic energy management.

The parallelization efficiency remains high even with increasing data size, indicating good scalability of the algorithm. This is particularly relevant in the context of increasing data volume in the energy sector.

The ability to efficiently process large volumes of data opens up new possibilities for energy management and evidence-based policy making, potentially transforming energy planning in Puno.

Improved accessibility to complex analyses can democratize decision making in the energy sector, allowing broader participation of local stakeholders.

Future work could explore:

The integration of deep learning techniques with ARIMA in a parallel context, potentially further improving prediction accuracy.

The implementation of this approach in distributed computing systems to handle even larger volumes of data, relevant for analyses at a national or broader regional level.

The application of this approach to other domains involving large-scale time series analysis, such as prediction of weather patterns or economic trends.

The development of tools and frameworks that facilitate the implementation of these parallel methods for researchers and practitioners not specialized in parallel computing, encouraging wider adoption.

The exploration of interpretability and explainability techniques for complex models, ensuring that predictions are understandable and reliable for decision makers.

In conclusion, parallel computing not only improves computational performance in predicting energy consumption, but can also be a catalyst for significant advances in energy management and planning in developing regions such as Puno, Peru. This approach has the potential to transform the way decisions are made in the energy sector, leading to more efficient and sustainable use of energy resources.

\end{document}